\documentclass[reprint,twoside,twocolumn,nofootinbib,showkeys,showpacs]{revtex4-1}% Specifies the document class %,unsortedaddress%
\usepackage{dcolumn}
\usepackage{latexsym}
\usepackage{amsmath}
\usepackage{amssymb}
\usepackage{amsthm}
\usepackage{dsfont}
\usepackage{longtable}
\usepackage{color}
\usepackage{verbatim}
\usepackage{graphicx}
\usepackage{epstopdf}
\usepackage{ulem}
\usepackage{cancel}

\usepackage[pdfstartview={FitH top},breaklinks]{hyperref} %,pdfpagemode=None
\hypersetup{
    colorlinks,
    citecolor=black,
    filecolor=black,
    linkcolor=black,
    urlcolor=blue
}
\theoremstyle{definition}

\makeatletter
\@addtoreset{remark}{utv}
\@addtoreset{remark}{theorem}
\@addtoreset{paragraph}{section}
\makeatother

\allowdisplaybreaks
\begin{document}
\title[]
{Correlation function intercepts for $\tilde{\mu},q$-deformed Bose gas model
implying effective accounting for interaction and compositeness of particles}

\author{A.~M. Gavrilik}%1
\affiliation{Bogolyubov Institute for Theoretical Physics of NAS of Ukraine,
14b Metrolohichna str., Kyiv 03680, Ukraine}
\email{omgavr@bitp.kiev.ua}%e-mail 1
\author{Yu.~A. Mishchenko}%2
\affiliation{Bogolyubov Institute for Theoretical Physics of NAS of Ukraine,
14b Metrolohichna str., Kyiv 03680, Ukraine}

%\address{}
\pacs{05.30.Jp, 05.90.+m, 11.10.Lm, 25.75.Gz} %\razd{\secix}
\keywords{Deformed Bose gas model, deformed bosons, deformation structure function,
two-particle distribution and correlation function intercept, two-pion
correlations at RHIC/STAR.}

\begin{abstract}
In the recently proposed two-parameter $\tilde{\mu},q$-deformed Bose
gas model [Ukr. J. Phys. {\bf 58}, 1171 (2013), arXiv:1312.1573] aimed to take
effectively into account both compositeness of particles and their
interaction, the $\tilde{\mu},q$-deformed virial expansion of the
equation of state (EOS) was obtained. In this paper we further
explore the $\tilde{\mu},q$-deformation, namely the version of
$\tilde{\mu},q$-Bose gas model involving deformed distributions and
correlation functions. In the model, we explicitly derive the one-
and two-particle deformed distribution functions and the intercept
of two-particle momentum correlation function. The results are
illustrated by plots, and the comparison with known experimental data on
two-pion correlation function intercepts extracted in RHIC/STAR
experiments is given.
\end{abstract}
\maketitle

\section{Introduction}

Deformed Bose gas models based on a set of identical deformed
(nonlinear) oscillators, or on deformed thermodynamic relations
provide nonlinear extension of standard Bose gas model which finds
applications to physical systems with one or more factors of
non-ideality~\cite{Avan,Scarfone2009Int,GM_Virial,GM2014VirCoefs,Rovenchak2014Polychr}.
In general, the effective description or modeling of essentially
nonideal (nonlinear) systems usually is performed by means of
reexpressing of the physical quantities of the initial complicated system
in terms of the analogous quantities of deformed model. Such two factors as
composite structure of particles of a gas and the interaction
between them are of main interest for us here. Concerning the
compositeness of particles, let us mention the
works~\cite{Avan,Bonatsos1992cor_pairs,Sviratcheva,Liu} where
$q$-deformed oscillators were applied for effective description of
composite particles (like nuclei, nucleons, mesons, excitons, cooperons,
atoms, molecules). Their Bose-Einstein condensation was also studied~\cite{Avancini_2003}.
It was shown in~\cite{GKM2,GKM} that two-fermionic (and two-bosonic)
composite bosons can be algebraically realized on their Fock
states by deformed oscillator algebra with the quadratically
nonlinear deformation structure function (DSF)
$\varphi_{\tilde{\mu}}(\hat{N}) = (1+\tilde{\mu})\hat{N} - \tilde{\mu} \hat{N}^2$,
with $\hat{N}$ the number operator, and discrete deformation parameter
$\tilde{\mu}=1/m$ involved. On the other hand, $q$-deformation of Arik-Coon
type~\cite{ArikCoon} based on the DSF equal to the ``$q$-bracket''
$[N]_q=\frac{q^N-1}{q-1}$ was used for the effective description~\cite{Scarfone_Interact}
of thermodynamics aspects (e.g., virial expansion) of Bose gas with
interaction.
So far, two aspects were treated {\it separately}, adhering to
different methods and contexts.  However, the task naturally arises of
treating {\it jointly}: {\it i}) compositeness of particles linked,
through the realization, with quadratic or
$\tilde{\mu}$-deformation; {\it ii}) the interaction between
particles modeled by $q$-deformation.  We may expect that combining
these two types of deformation into single one will reproduce
effectively, in a unified manner, some specific features inherent to
the thermodynamic or statistical quantities of more realistic
systems of particles possessing {\it both interaction and
compositeness}.  The simplest variant of such unification is their
functional composition or $\tilde{\mu},q$-deformation. Of course, at
the moment the ascription of the meaning of deformation parameters
$\tilde{\mu}$ and $q$ as responsible respectively for the
compositeness and the interaction is rather formal, and the detailed
consistent analysis providing a reformulation in the deformed model
terms, including the relation with the parameters of interaction or
compositeness, is not completed to sufficient extent.

First steps to the (microscopics of) effective taking of interaction
and compositeness jointly into account are made by introducing the
$\tilde{\mu},q$-deformed Bose gas model~\cite{GM_Virial,GM2014VirCoefs} based on
deforming the thermodynamics.
Namely, in~\cite{GM_Virial} the $\tilde{\mu},q$-deformed Bose gas model was realized
through deforming the total mean number of particles or the partition function by
means of the deformed analog of the derivative $z\frac{d}{dz}$ ($z$ is
fugacity) and use of the ``hybrid'' (combined) DSF
\begin{equation}\label{phi_muq}
\varphi_{\tilde{\mu},q}\Bigl(z\frac{d}{dz}\Bigr)\!\!\equiv\!
\varphi_{\tilde{\mu}}(D_q) \!=\! (1\!+\tilde{\mu})D_q -
\tilde{\mu} D_q^2,\  D_q\!\equiv\!\!
\Bigl[z\frac{d}{dz}\Bigr]_q\!.\!
\end{equation}
This version of $\tilde{\mu},q$-Bose gas model was constructed
bearing in mind the goal of effective description of thermodynamics
of {\it interacting}  gas of {\it composite} bosons (made of two
bosons or two fermions). In fact, the latter due to compositeness are
no longer true bosons. For that model, the deformed virial
expansion was studied. In the sequel to ~\cite{GM_Virial}, the
relation of the obtained virial coefficients of the
$(\tilde{\mu},q)$-deformed model (dependent explicitly on the deformation
parameters $\tilde{\mu}$ and $q$) with scattering data of some
interaction was explored~\cite{GM2014VirCoefs}, and the arising
unusual temperature dependence of $\tilde{\mu}$ and $q$ discussed and justified.

The version of $\tilde{\mu},q$-deformed Bose gas model considered
herein uses for its definition the same DSF
$\varphi_{\tilde{\mu},q}$ as in~\cite{GM_Virial} (composed of
quadratically nonlinear SF and the $q$-deformed one). However, this
time the DSF is exploited as the operator function of $\hat{N}$ for
obtaining one- and two-particle distribution functions. The
resulting model is also called $(\tilde{\mu},q)$-deformed model.

Below we focus on one- and two-particle distributions and the
related momentum correlation function intercept, and calculate them.
Let us quote some preceding
activity~\cite{Man'ko1993Corr,DaoudKibler1995,Anchishkin,ZhangPadula2004,Adamska,GR_EPJA,GM_Exact,GM2014STD}
on the (intercepts of) correlation functions for various deformed
Bose gas models. The knowledge of the correlation function intercept is
useful for an application in the effective modeling of non-Bose like
properties of data on pionic intercepts extracted in RHIC and LHC
experiments~\cite{STAR2001pion,Bearden(CERN)2001,STAR2003HBTCorr,Aggarwal(WA98)2003,STAR2005Pion,STAR_pion2009,ALICE2014pion}.
That is, as a particular physical system for which the considered
$\tilde{\mu},q$-deformed intercepts can be applied we mean the
$\pi$-mesonic gas created in relativistic heavy-ion collisions.
Obviously this is the gas of composites (quark-antiquark bound
states) which moreover undergo interaction. Remark however that there are
some further complications: {\it i)} this is non-equilibrium system,
{\it ii)} the mentioned ``interaction'' may not reduce to only
simple $\pi+\pi$-interaction, and {\it iii)} the non-ideality
factors may play the prevailing role at the stage of $\pi$-meson
formation (memory effects etc.). Nevertheless, for the results of the studied model we
will make a comparison with some of the available experimental data on
$\pi$-meson correlation function intercepts. Note that, using other
deformed Bose gas models, the comparison with experimental data on
the $\pi$-mesonic correlation function intercepts of 2nd (and 3rd)
order was considered in some earlier papers,
e.g. in~\cite{Anchishkin_Transverse,Gavrilik_Sigma,GR_EPJA}.

\section{Deformed Bose gas model with $\tilde{\mu},q$-deformed partition function:
virial expansion of EOS} \label{sec:vir_exp}

In this section we give an overview of the results from~\cite{GM_Virial}
where basing on~\cite{GKM,GKM2} and on \cite{Scarfone_Interact} the
specially designed two-parameter $\tilde{\mu},q$-deformed Bose gas
model capable to effectively describe the {\it interacting} gas of
{\it composite} bosons was constructed. As we are interested in the
effective description of interaction and compositeness effects
existing in realistic gases, the model from~\cite{GM_Virial,GM2014VirCoefs}
can serve as the ``link'' to the study of more microscopical aspects
(including the involved parameters), especially in view of~\cite{GM2014VirCoefs}.
%and the following relation. The two mentioned types of models (the one from~\cite{GM_Virial,GM2014VirCoefs}
%on the one hand, and the model this paper is devoted to, defined by deformed distributions,
%on the other hand) have equal one-particle distributions, and thus the similar behavior in
%the corresponding sector of physical quantities, if the DSFs defining them satisfy certain
%relation (are interrelated)~\cite{GM_duality}, $\tilde{\varphi}_{\tilde{\mu},q} = \tilde{\varphi}(\varphi_{\tilde{\mu},q})$.}
As mentioned, the corresponding DSF $\varphi_{\tilde{\mu},q}$ in (\ref{phi_muq})
which determines the deformed Bose gas model to be considered in
this section is the combination of previously studied ones,
see~\cite{Scarfone_Interact} and the works~\cite{GKM,GKM2}.
For the $\tilde{\mu},q$-deformed Bose gas model~\cite{GM_Virial},
the corresponding deformed virial expansion was obtained along with
first five virial coefficients, and interpreted as the virial
expansion accounting for both the interaction of (composite) bosons
and the very their compositeness. The thermodynamic relations for
the deformed Bose gas model including partition function and the
equation of state (EOS), which were used in the process of
derivation of its virial expansion, were obtained by using the
$\tilde{\mu},q$-generalization (\ref{phi_muq}) of the Jackson
derivative, adjusted for the concerned unifying deformation (note
that in~\cite{GKR_UJP2013} similar procedure of deformation was
applied within differently motivated deformed model, the $\mu$-Bose gas model).

\paragraph{\bf Compositeness aspects.}

The creation and annihilation operators for composite bosons
in the second quantization scheme are constructed \cite{Avan,GKM2,Tichy2013Two-boson} %\my{/\cite{Tichy2012Bos_beh}}
from two-fermion (or two-boson) operators as
\[
A^\dag_\alpha = \sum_{\mu\nu} \Phi_\alpha^{\mu\nu} a^\dag_\mu
b^\dag_\nu,\quad A_\alpha = \sum_{\mu\nu}
\overline{\Phi_\alpha^{\mu\nu}} b_\nu a_\mu.
\]
Here $a^\dag_\mu$, $b^\dag_\nu,$ and $a_\mu$, $b_\nu$ are the
creation and annihilation operators for the constituents; the
matrices $\Phi_\alpha^{\mu\nu}$ determine the composite boson
wavefunction. The operators $A_\alpha$ and $A^\dag_\beta$ obey the
relation $[A_\alpha,A^\dag_\beta] = \delta_{\alpha\beta} -
\Delta_{\alpha\beta}$ with
\[
\Delta_{\alpha\beta} =
\sum_{\mu\mu'}(\Phi_{\beta}{\Phi}^{\dag}_{\alpha})^{\mu'\mu}
a^{\dag}_{\mu'}a_{\mu} + \sum_{\nu\nu'}({\Phi}^{\dag}_{\alpha}
\Phi_{\beta})^{\nu\nu'} b^{\dag}_{\nu'}b_{\nu}
\]
($\Delta_{\alpha\beta}$ reflects a deviation from pure bosonic case).

The many-body system of composite (two-fermion or two-boson) quasi-bosons with certain
composite wave function can be realized at the operator level, see~\cite{GKM,GKM2}, by a
deformed Bose gas model with quadratic DSF ($\tilde{\mu}\ge 0$)
\begin{equation}\label{phi_mu}
\varphi_{\tilde{\mu}}(N)\!\equiv\! [N]_{\tilde{\mu}} \!=\!
\left\{
\begin{aligned}
&(1\!+\!\tilde{\mu})N \!-\! \tilde{\mu} N^2\ \text{(two-fermion)},\\
&(1\!-\!\tilde{\mu})N \!+\! \tilde{\mu} N^2\ \text{(two-boson)},
\end{aligned}
\right.
\end{equation}
involving the discrete deformation parameter $\tilde{\mu}=1/m$, $m=1,2,...$\,.
Then, the gas of composite bosons can be treated (at least on the states) as
the corresponding gas of deformed bosons. Note that such a realization of composite bosons
by deformed oscillators is of importance in quantum information theory, as it was
demonstrated in~\cite{GM_Entang,GM_Ent(En)} where the characteristics of
bipartite entanglement were expressed directly through the deformation parameter $\tilde{\mu}$.

\paragraph{\bf Deformed Bose gas accounting for compositeness of
particles.} \label{sec:compos}

So, the $\tilde{\mu}$-deformed bosons with the quadratic DSF of the form (\ref{phi_mu})
do realize~\cite{GKM,GKM2} the two-fermion (or two-boson) composite Bose-like particles.
For the deformed thermodynamics of $\tilde{\mu}$-Bose gas, the deformed virial expansion
of the EOS has been derived~\cite{GM_Virial} along with the first five virial coefficients,
by using the $\tilde{\mu}$-deformed derivative (compare with (\ref{phi_mu})):
\[
\mathcal{D}_z^{(\tilde{\mu})} z^n = ((1+\varkappa\tilde{\mu})n -
\varkappa\tilde{\mu} n^2) z^{n-1}, \quad \varkappa=\pm1.
\]

In the $\tilde{\mu}$-deformed picture, similarly to~\cite{GKR_UJP2013},
we obtain the mean number of particles $N^{(\tilde{\mu})}$ (with $Z$
denoting non-deformed partition function) as
\begin{equation}
N^{(\tilde{\mu})} \!=\! \Bigl[z\frac{d}{dz}\Bigr]_{\tilde{\mu}}\!
\ln Z \!\equiv\! z \mathcal{D}_z^{(\tilde{\mu})}\! \ln Z \!=\!
\frac{V}{\lambda^3}\! \sum_{n=1}^\infty [n]_{\tilde{\mu}}
\frac{z^n}{n^{5/2}}\label{N^mu}
\end{equation}
or, using the notation $v=\frac{V}{N^{(\tilde{\mu})}}$, as
\begin{equation}\label{lambda/v}
\frac{\lambda^3}{v} = \sum_{n=1}^\infty [n]_{\tilde{\mu}}
\frac{z^n}{n^{5/2}}.
\end{equation}
In~(\ref{N^mu}), (\ref{lambda/v}) and below, the $\tilde{\mu}$-bracket means:
$[X]_{\tilde{\mu}}\equiv(1+\tilde{\mu})X -\tilde{\mu}X^2$.

The $\tilde{\mu}$-deformed partition function $Z^{(\tilde{\mu})}$
is then obtained from (\ref{N^mu}) by applying the inversion:
\begin{equation}
\ln Z^{(\tilde{\mu})} = \Bigl(z\frac{d}{dz}\Bigr)^{\!-1} N^{(\tilde{\mu})} =
\Bigl(\frac{d}{dz}\Bigr)^{\!-1} \mathcal{D}_z^{(\tilde{\mu})} \ln Z.
\end{equation}
As result, the deformed EOS takes the form
\begin{multline}\label{eq_st(z)}
\frac{PV}{k_{\rm  B} T} = \ln Z^{(\tilde{\mu})} =\\
= \frac{V}{\lambda^3}
\Bigl(\!z\!+\!\frac{[2]_{\tilde{\mu}}}{2^{7/2}}z^2\!+\!\frac{[3]_{\tilde{\mu}}}{3^{7/2}}z^3\!+\!
\frac{[4]_{\tilde{\mu}}}{4^{7/2}}z^4\!+\!\frac{[5]_{\tilde{\mu}}}{5^{7/2}}z^5+...\Bigr).
\end{multline}
More informative is the virial expansion of EOS that involves the
series in powers of $\frac{\lambda^3}{v}$. The desired virial
expansion~\cite{GM_Virial} modeling that of (non-interacting) gas of
{\it composite bosons} and depending on the parameter $\tilde{\mu}$ reads
\begin{multline}\label{Virial_mu}
\frac{P v}{k_{\rm  B} T} = \sum_{k=1}^\infty
V_k(\tilde{\mu}) \Bigl(\frac{\lambda^3}{v}\Bigr)^{k-1} =\\
= 1 - \frac{[2]_{\tilde{\mu}}}{2^{7/2}}
\frac{\lambda^3}{v} + \Bigl(\frac{[2]^2_{\tilde{\mu}}}{2^5} -
\frac{2[3]_{\tilde{\mu}}}{3^{7/2}}\Bigr)\Bigl(\frac{\lambda^3}{v}\Bigr)^{\!2}+\\
+ \Bigl(-\frac{3[4]_{\tilde{\mu}}}{4^{7/2}} +
\frac{[2]_{\tilde{\mu}}[3]_{\tilde{\mu}}}{2^{5/2}3^{3/2}} -
\frac{5[2]^3_{\tilde{\mu}}}{2^{17/2}}\Bigr)\Bigl(\frac{\lambda^3}{v}\Bigr)^{\!3}
+...
%+ \Bigl(-\frac{4[5]_{\tilde{\mu}}}{5^{7/2}} +
%\frac{[2]_{\tilde{\mu}}[4]_{\tilde{\mu}}}{2^{11/2}} -
%\frac{2[3]^3_{\tilde{\mu}}}{3^5} -
%\frac{[2]^2_{\tilde{\mu}}[3]_{\tilde{\mu}}}{2^3 3^{3/2}} +
%\frac{7[2]^4_{\tilde{\mu}}}{2^{10}}\Bigr)\times\\
% \times
%\Bigl(\frac{\lambda^3}{v}\Bigr)^{\!4} + ...\Bigr\}.
\end{multline}
Since the second virial coefficient $V_2 =
-\frac{[2]_{\tilde{\mu}}}{2^{7/2}} = -
\frac{1-\tilde{\mu}}{2^{5/2}}$ vanishes at $\tilde{\mu}\to 1$,
such a nullifying can be interpreted as (due to) mutual
compensation, at $\tilde{\mu}=1$, of the compositeness effects
against the quantum-statistical many-particle effects, so that the
quantum gas of composite bosons then behaves like a classical gas of
pointlike particles, at least to the first order in $\lambda^3/v$.
This fact, on the other hand, implies that the compositeness effects
measured by $\tilde{\mu}$ can be interpreted as another amount
of effective interaction (between quasibosons) contributing to $V_2$
similarly, though with opposite sign, to the bosonic quantum
statistical many-particle effects. Similar analysis may be applied
to higher virial coefficients $V_3$, $V_4$, etc.

\paragraph{\bf Account for the interaction between (elementary) Bose particles.}
%\\ Described by a \boldmath$q$-Deformed Algebra}\label{sec:interact}

The interpretation~\cite{Scarfone_Interact} of interacting
many-boson systems in terms of $q$-deformed oscillators ($q$-bosons)
is based on the assumption that a suitably chosen $q$-deformed
thermodynamic or statistical relation for non-interacting
structureless system can be applied, within  some approximation, to
model interacting many-boson system with certain interaction.
In this sense, an effective description of the interacting gas of
Bose particles was dealt with in~\cite{Scarfone_Interact} by means
of $q$-deformation with the structure function $\varphi_q(n)\equiv
[n]_q$. The effects of interaction between particles
of the Bose gas were incorporated in such deformed model
by means of $q$-deformed thermodynamic relations.  For instance, the respective specific
volume (as function of fugacity) was obtained in its $q$-deformed version.

Using the series expansion of the basic-number like operator $[N]_q$
in terms of $\epsilon\equiv q-1$, natural interpretation is got as
the picture of incorporating the interparticle interaction, since
the contributions due to interaction can be viewed either in terms
of $N$, $N^2$, $N^3$,~... or in terms of $\epsilon$, $\epsilon^2$,
$\epsilon^3$, ...\,. In both cases the parameters characterizing
interaction enter the terms (coefficients) depending on the
deformation parameter $q$ or $\varepsilon=q-1$.

The Hamiltonian for $q$-boson is taken as $H_\epsilon(N)$ being some
$\epsilon$-deformation of standard quantum oscillator Hamiltonian.
The expansion of $H_\epsilon$ in powers of $\epsilon$ is interpreted
in a similar fashion: the terms of the first and higher orders in
$\epsilon$ in the expansion are again viewed as those linked with
interaction. That is, they alltogether constitute the interaction
Hamiltonian, and this implies physical meaning.

So, the picture of $q$-deformed non-interacting (ideal)
many-particle system is used as a model of non-deformed, but
interacting system. The $q$-deformed virial expansion is written in
the form~\cite{Scarfone_Interact}:
\begin{equation}
\frac{Pv}{k_{\rm  B} T} = \sum_{k=1}^\infty a_k(\epsilon)
\Bigl(\frac{\lambda^3}{v}\Bigr)^{k-1},
\end{equation}
with the virial coefficients $a_k(\epsilon)$ given, say up to $\epsilon^3$, as
$\ a_2(\epsilon) = -\frac{1}{4\sqrt{2}}-\frac{1}{48\sqrt{2}} \epsilon^2(1-\epsilon)
+ O(\epsilon^4)$ and likewise for higher $a_n(\epsilon)$.
Similarly to the preceding interpretation, $\epsilon\ne0$ terms suggest corrections to the
standard virial coefficients of the ideal Bose gas viewed as those
arising from some explicitly given (though unspecified) interaction described by certain
potential in the Hamiltonian. In effect, the interacting many-particle system gets
effectively described (and interpreted) in terms of non-interacting, deformed system.

\paragraph{\bf Account for both compositeness and interaction of particles.}

Above, due to the realizability of composite bosons by deformed bosons, we deformed the
Bose gas model with (the quadratic) structure function $\varphi_{\tilde{\mu}}(n)$ in
order to find effective thermodynamic relations or functions for the ideal, non-interacting
{\it quantum gas of composite bosons}, in particular the deformed (i.e. depending on
$\tilde{\mu}$) virial expansion of the EOS. To take into account the interaction between
particles jointly with their compositeness, the two DSFs $\varphi_{\tilde{\mu}}(n)$
and $\varphi_q(n)$ are combined into a single one yielding the unified DSF
\begin{equation}\label{phi_mu_q}
\varphi_{\tilde{\mu},q}(n) = (1+\tilde{\mu})[n]_q - \tilde{\mu}
([n]_q)^2 \equiv [n]_{\tilde{\mu},q}
\end{equation}
which will play basic role in our treatment. For modeling the effects of
the interaction between particles jointly with their compositeness,
in parallel to DSF~(\ref{phi_mu_q}) the corresponding
$(\tilde{\mu},q)$-extension of the derivative was used,
\[
z \mathcal{D}_z^{(\tilde{\mu},q)} = (1+\tilde{\mu})
\Bigl[z\frac{d}{dz}\Bigr]_{q} - \tilde{\mu} \Bigl[z\frac{d}{dz}\Bigr]_{q}^2.
\]

The two-parameter Hamiltonian $H_{\tilde{\mu},\epsilon}(N)$ of
${\tilde{\mu},q}$-bosons (single-mode case) can be split into $H_0$ (non-deformed part)
and the Hamiltonian $H_1(\epsilon,\tilde{\mu}; N)$
that depends on $N$ and is the double series in $\tilde{\mu}$ and $\epsilon=q-1$.

Using DSF $\varphi_{\tilde{\mu},q}(n)$ and the $\tilde{\mu},q$-derivative
$\mathcal{D}_z^{(\tilde{\mu},q)}$ similarly to (\ref{lambda/v})-(\ref{eq_st(z)}),
the virial expansion of the EOS results in the form~\cite{GM_Virial}
\begin{multline}\label{Virial_mu_q}
\frac{Pv}{k_{\rm  B} T} =\sum_{k=1}^\infty
V_k(\tilde{\mu},q) \Bigl(\frac{\lambda^3}{v}\Bigr)^{k-1}=
1 - \frac{[2]_{\tilde{\mu},q}}{2^{7/2}} \frac{\lambda^3}{v} +\\
+\Bigl(\frac{[2]^2_{\tilde{\mu},q}}{2^5} -
\frac{2[3]_{\tilde{\mu},q}}{3^{7/2}}\Bigr)\Bigl(\frac{\lambda^3}{v}\Bigr)^{\!2}+\\
+ \Bigl(-\frac{3[4]_{\tilde{\mu},q}}{4^{7/2}} +
\frac{[2]_{\tilde{\mu},q}[3]_{\tilde{\mu},q}}{2^{5/2}3^{3/2}}
- \frac{5[2]^3_{\tilde{\mu},q}}{2^{17/2}}\Bigr)\Bigl(\frac{\lambda^3}{v}\Bigr)^{\!3} + ...\,.
%+ \Bigl(-\frac{4[5]_{\tilde{\mu},q}}{5^{7/2}} +
%\frac{[2]_{\tilde{\mu},q}[4]_{\tilde{\mu},q}}{2^{11/2}}-\\
%- \frac{2[3]^3_{\tilde{\mu},q}}{3^5} -
%\frac{[2]^2_{\tilde{\mu},q}[3]_{\tilde{\mu},q}}{2^3 3^{3/2}}+
%\frac{7[2]^4_{\tilde{\mu},q}}{2^{10}}\Bigr)\Bigl(\frac{\lambda^3}{v}\Bigr)^{\!4}
\end{multline}

It is tempting to interpret this virial expansion
as the effective one corresponding to the {\it interacting} gas of
{\it composite} particles. The information about interaction
and the composite structure is respectively encoded in the
deformation parameters $q$ and $\tilde{\mu}$.
 If $\tilde{\mu}=0$, $q\ne 1$, the expansion~(\ref{Virial_mu_q})
accounts solely for interaction between the particles; likewise,
when $q=1$, $\tilde{\mu}\ne 0$, formula (\ref{Virial_mu_q}) should
be interpreted as accounting for the compositeness of particles.
When both $q\ne1$ and $\tilde{\mu}\ne0$, expression (\ref{Virial_mu_q})
incorporates jointly the both mentioned factors of Bose gas non-ideality.

The explicit virial coefficients with their dependence on the deformation
parameters $q$ and $\tilde{\mu}$ can in principle be related to the characteristic
parameters linked directly/explicitly with the interaction between composite bosons as well as
inside them (between their constituents), see~\cite{GM2014VirCoefs}.

Remark that in~\cite{GM_Virial} alternative DSFs defining the deformed Bose
gas model suitable for the effective description were also discussed. Those correspond
to other ways of composing the $q$-deformed SF and the quadratic one, in particular
such as the DSF $\varphi_q(\varphi_{\tilde{\mu}}(n))$.
%-------------------------------------------------------------------------------------------------

\section{One- and two-particle distributions} \label{sec:2nd-distr}

Our main goal is to calculate the intercept of the momentum
correlation function of 2nd order for $\tilde{\mu},q$-deformed Bose
 gas model defined by the DSF $\varphi_{\tilde{\mu},q}(n)$ in~(\ref{phi_mu_q}),
%%(1+\tilde{\mu})[n]_q - \tilde{\mu} ([n]_q)^2$
 with the above-given interpretation (when $\tilde{\mu}$ and $q$ are
 responsible resp. for the compositeness and interaction) this time without
 the appeal to virial expansions.

We start with the defining formula (see e.g.~\cite{ChapmanHeinz1994}
for nondeformed case, and~\cite{Adamska} for deformed one)
\begin{equation}\label{lambda_def}
\lambda^{(r)}({\bf k}) = \frac{\langle (a^\dag_{\bf k})^r (a_{\bf k})^r \rangle}{\langle a^\dag_{\bf k} a_{\bf k}\rangle^r}-1
\end{equation}
for the intercepts of $r$th order momentum correlation functions at a
given momentum~$\bf k$, taken the same for all the $r$ particles. The
notation $\langle...\rangle$ means statistical (thermal) average,
and $a^\dag_{\bf k}$ resp. $a_{\bf k}$ are the creation resp.
annihilation operators for the $\varphi_{\tilde{\mu},q}$-deformed
bosons which obey the following set of commutation relations given
by the DSF $\tilde{\varphi}_{\tilde{\mu},q}$:
\begin{align*}
&[N_{\bf k},a^\dag_{\bf k'}] = \delta_{\bf kk'} a^\dag_{\bf k},\quad
[N_{\bf k},a_{\bf k'}] = -\delta_{\bf kk'} a_{\bf k},\\
&[a_{\bf k},a^\dag_{\bf k}] = \varphi_{\tilde{\mu},q}(N_{\bf k}+1) - \varphi_{\tilde{\mu},q}(N_{\bf k}),\quad
a^\dag_{\bf k}a_{\bf k} = \varphi_{\tilde{\mu},q}(N_{\bf k}).
\end{align*}
From these relations it follows that
(from now on the label $\bf k$ of fixed mode will be omitted)
\begin{equation}\label{a_phi}
a^\dag \varphi_{\!\tilde{\mu},q}(N) \!=\! \varphi_{\!\tilde{\mu},q}(N\!-1) a^\dag,\ \,
a \varphi_{\!\tilde{\mu},q}(N) \!=\! \varphi_{\!\tilde{\mu},q}(N\!+1) a.
\end{equation}
As seen, (\ref{lambda_def}) involves both the $r$-th order (in
numerator) and $r$th power of the first order (in denominator)
deformed analogs of distribution functions.

\begin{figure}[h]
\centering
\includegraphics[width = 1\linewidth]{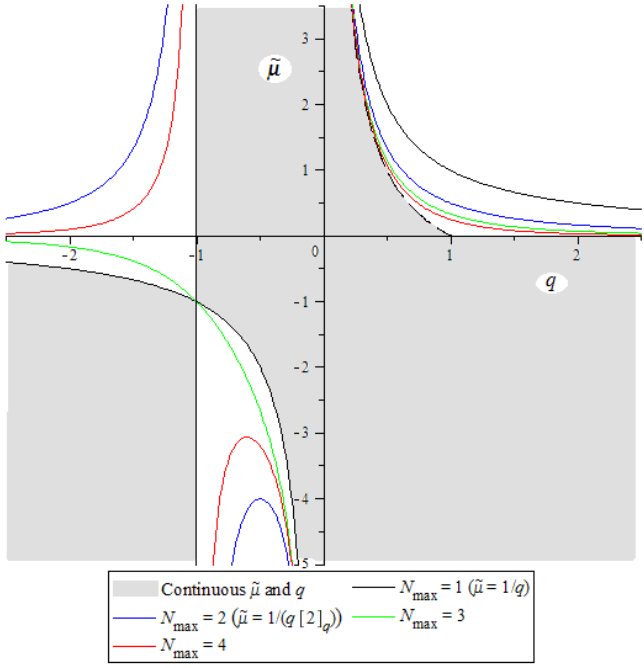}
\caption{Admissible deformation parameters $\tilde{\mu}$, $q$, and the
maximum occupation number $N_{\rm max}$ (discrete subset).}
\label{fig1}
\end{figure}

\begin{figure}[h]
\centering
\includegraphics[width = 1\linewidth]{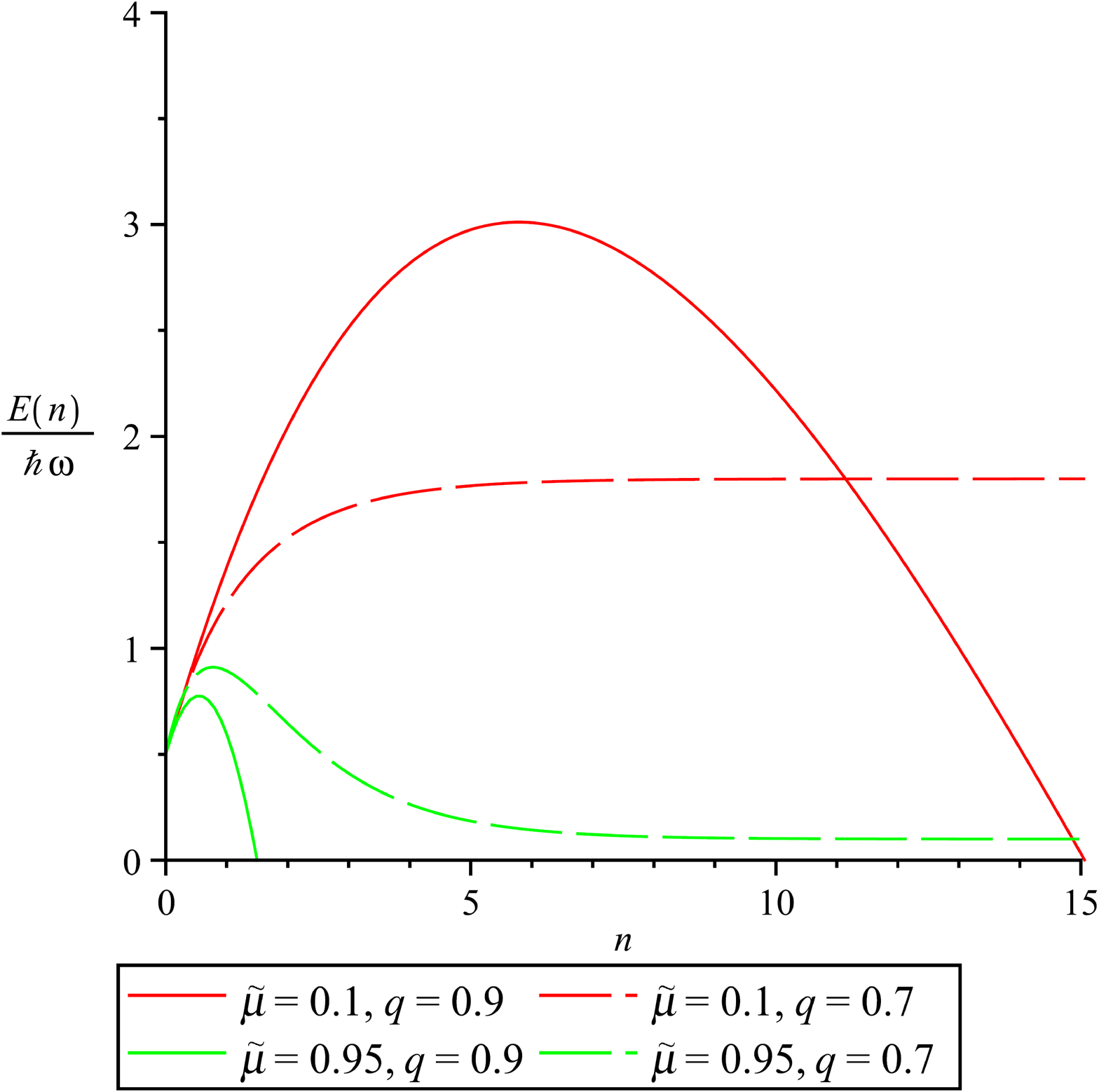}
\caption{Energy $E(n)=\frac12 \hbar\omega (\varphi_{\tilde{\mu},q}(n)+\varphi_{\tilde{\mu},q}(n+1))$
versus occupation number $n$, for some values of deformation parameters $\tilde{\mu}$ and $q$.}
\label{fig5}
\end{figure}

Let us observe the nilpotency of creation/annihilation operators $a^\dag$,
$a$ for certain $\tilde{\mu},\,q$, and, as a consequence, the partially
discontinuous (either in $\tilde{\mu}$ or in $q$) set of deformation
parameters~$(\tilde{\mu},\,q)$ for $\tilde{\mu},q$-deformed oscillators.
The nilpotency of $a^\dag$, $a$ is related to the possibility of
nullifying or changing the sign of the structure function
$\varphi_{\tilde{\mu},q}(n)$ at some positive $n$. The reasoning
is as follows. The norm of the vector $(a^\dag)^r|0\rangle$ squared, i.e.
\begin{multline}
\|(a^\dag)^r|0\rangle\|^2 = \langle 0|a^r(a^\dag)^r|0\rangle =\\
= \varphi(r)\varphi(r-1)\cdot...\cdot\varphi(1)\equiv \varphi(r)!
\end{multline}
should be nonnegative: either positive for all $r\ge1$ (unbounded occupation
numbers) or zero at some $r=N_{\rm max}+1$ where $N_{\rm max}$ is maximal
occupation number. The former requirement, for the DSF
$\varphi_{\tilde{\mu},q}(n)$, is equivalent to the condition
\begin{equation}\label{cont_reg}
\mathop{\min}\limits_{n\ge1} \varphi_{\tilde{\mu},q}(n) > 0 \ \ \text{or}\ \
\varphi_{\tilde{\mu},q}(2) > [2]_q \frac{1\!+\!|q|\!-\!|q\!-\!1|}{2}.
\end{equation}
So, the set of parameters $(\tilde{\mu},q)$ is continuous inside the
two-dimensional region given by inequalities in~(\ref{cont_reg})
(grey-colored region in Fig.~\ref{fig1}). Otherwise, there should exist an integer $r=N_{\rm
max}+1$ such that
\begin{equation}\label{phi_N_max}
\varphi_{\tilde{\mu},q}(r)=\varphi_{\tilde{\mu},q}(N_{\rm max}+1)=0.
\end{equation}
This equation can be equivalently rewritten as
\begin{equation}\label{q^N_max}
[N_{\rm max}+1]_q=0\ \ \ {\rm or}\ \ \ \tilde{\mu} q [N_{\rm max}]_q = 1.
\end{equation}
Its solutions $(\tilde{\mu},q)$ form a discrete set of curves, see
Fig.~\ref{fig1}, as they are parameterized by the integer $N_{\rm
max}$. Thus, the possibilities of discrete $\tilde{\mu} = q^{-1}
[N_{\rm max}]_q^{-1}$ in couple with continuous $q$, or of discrete
$q=q(\tilde{\mu},N_{\rm max})$ with continuous $\tilde{\mu}$ are
included. However, the former seems more probable in view of
discreteness of the set of composite bosons' bound states. Note that
other intermediate variants, e.g. when a certain function of $\tilde{\mu}$,
$q$ is continuous are also allowed. From~(\ref{q^N_max})
we deduce the maximal occupation number of a fixed mode for the
$\tilde{\mu},q$-deformed Bose gas model with $\tilde{\mu},q$
belonging to the discrete curves:
\begin{equation}\label{N_max}
N_{\rm max}(\tilde{\mu},q) = \ln\Bigl(1 + {\frac{q-1}{\tilde{\mu} q}}\Bigr)/\ln q.
\end{equation}
Remark that at fixed values of $q$ the corresponding set of
deformation parameter $\tilde{\mu}$ values (discrete ones plus
regions of continuum), see Fig.~\ref{fig1}, can be associated with
bound and unbound states. The dependence of energy $E(n)$ on the
occupation number $n$ can be of interest for diverse values of the
deformation parameters.  The respective dependences for the typical
Hamiltonian of deformed oscillator $H=\frac12 \hbar\omega
(aa^\dag+a^\dag a)$, for which $E(n)=\frac12 \hbar\omega
(\varphi_{\tilde{\mu},q}(n+1)+\varphi_{\tilde{\mu},q}(n))$, are
shown in Fig.~\ref{fig5}.  Let us notice two different types of
behavior of $E(n)$ as function of $n$: for the first one there
exists such $n$ that  $E(n)$ becomes non-positive, and for the
second one $E(n)>0$ for all $n$.  This is linked with two mentioned
types of behavior, i.e. with the presence of maximal occupation
number $N_{\rm max}(\tilde{\mu},q)$ determined in~(\ref{N_max}). The
first (discrete) type holds when $q [N_{\rm max}]_q = 1$,
$N_{\rm max}\ge 1$ or when $q=-1$, $\tilde{\mu}$ is arbitrary, $N_{\rm max}=1$;
the second (continuous) type holds when
$\varphi_{\tilde{\mu},q}(2) > [2]_q \frac{1+|q|-|q-1|}{2}$.

For the gas of $\tilde{\mu},q$-bosons we take the simplest many-particle system Hamiltonian
\begin{equation}\label{H}
H= \sum_{\bf k} \hbar\omega_{\bf k} N_{\bf k}.
\end{equation}
Then for the average $\langle a^\dag_{\bf k} a_{\bf k}\rangle$
we obtain ($\bf k$th mode is fixed; its label $\bf k$ is dropped)
\begin{multline}\label{<a^+a>2}
\langle a^\dag a\rangle \!=\! \frac{1}{z\!-\!q} \!+\! \frac{(\varphi_{\tilde{\mu},q}(2)\!-\![2]_q)
\Bigl(1\!-\!\frac{[N_{\rm max}\!+1]_{q^2}}{[N_{\rm max}\!+1]_z}\Bigr)}{(z\!-\!q)(z\!-\!q^2)}=\\
= \frac{1}{z\!-\!q} \!+\! \frac{\delta\varphi_{\tilde{\mu},q}(2) (1\!-\!R)}{(z\!-\!q)(z\!-\!q^2)},
\end{multline}
where
\begin{equation*}
\delta\varphi_{\tilde{\mu},q}(n)\!\equiv\! \varphi_{\tilde{\mu},q}(n)\!-\![n]_q,
\ \ \ R\!\equiv\! R_{\tilde{\mu},q}(z)\!=\!\frac{[N_{\rm max}\!+1]_{q^2}}{[N_{\rm max}\!+1]_z},
\end{equation*}
$z\!=\!e^x$, $x\!=\!\beta \hbar\omega$, $\beta\!=\!(k_B T)^{-1}$, $k_B$ is Boltzmann
constant, in the discrete case, i.e. when $\tilde{\mu} q [N_{\rm max}]_q = 1$, and
\begin{equation}
\langle a^\dag a\rangle = \frac{z+\varphi_{\tilde{\mu},q}(2)-[3]_q}{(z-q)(z-q^2)}
\end{equation}
in the continuous case when $\varphi_{\tilde{\mu},q}(2) > [2]_q \frac{1+|q|-|q-1|}{2}$.
{\it This is our first result.}

Let us consider the $q\to 1$ and $\tilde{\mu}\to 0$ limits of $\langle a^\dag a\rangle$.
The evaluation yields
\begin{equation}
\langle a^\dag a\rangle \mathop{\longrightarrow}\limits_{q\to 1}
\left\{
\begin{aligned}
&\frac{z\!-\!1\!-\!2\tilde{\mu}}{(z-1)^2} + 2\frac{1\!+\!N_{\rm max}^{-1}}{[N_{\rm max}\!+1]_z}\frac{1}{(z\!-\!1)^2},
\ \ \tilde{\mu} \!=\! N_{\rm max}^{-1},\\
&\frac{z\!-\!1\!-\!2\tilde{\mu}}{(z-1)^2},\ \ \tilde{\mu}<0.
\end{aligned}
\right.
\end{equation}
On the other hand, for $\tilde{\mu}\to 0$ and $q>1$ we have:
\begin{align}
\langle a^\dag a\rangle_{\tilde{\mu}\to 0} &\simeq %% ??????
\frac{1}{e^x-q} + \theta(\tilde{\mu}) q \frac{e^x-1}{(e^x\!-\!q)(e^x\!-\!q^2)}
\Bigl|\frac{\tilde{\mu}}{q-1}\Bigr|^{\frac{x}{\ln q}-1}\nonumber\\
&\mathop{\longrightarrow}\limits_{\tilde{\mu}\to 0}
\left\{
\begin{aligned}
&\frac{1}{e^x-q},\ \ \ x>\ln q;\\
&\infty,\ \ \ \ x<\ln q,
\end{aligned}
\right.
\end{align}
where $\theta(\tilde{\mu})$ is the Heaviside step function.
Otherwise for $|q|<1$ the latter limit yields
\begin{equation}
\langle a^\dag a\rangle \mathop{\longrightarrow}\limits_{\tilde{\mu}\to 0} \frac{1}{e^x-q},
\end{equation}
which is the familiar result for Arik-Coon type $q$-Bose gas (see e.g.~\cite{Adamska}).

With account of the easily verified equality
\begin{equation}
(a^\dag)^r a^r = \varphi(N)\varphi(N-1)\cdot...\cdot\varphi(N-r+1)
\end{equation}
(see~(\ref{a_phi})) we consider the 2-particle $\tilde{\mu},q$-deformed
distribution function: $\langle (a^\dag)^2 a^2
\rangle_{\tilde{\mu},q} = \langle
\varphi_{\tilde{\mu},q}(N)\varphi_{\tilde{\mu},q}(N-1)\rangle$.
To calculate $\langle (a^\dag)^2 a^2 \rangle_{\tilde{\mu},q}$ we use the
relation
\begin{multline}\label{recur2}
\sum\nolimits_{i=0}^4 (-1)^i q^{i(i+1)/2} \Bigl({\scriptstyle 4\atop\scriptstyle i}\Bigr)_q
\varphi_{\tilde{\mu},q}(N\!+\!4\!-\!i)\varphi_{\tilde{\mu},q}(N\!+\!3\!-\!i) = \\
=\sum\nolimits_{i=0}^2 (-1)^i q^{i(i+1)/2} \Bigl({\scriptstyle 4\atop\scriptstyle i}\Bigr)_q
\varphi_{\tilde{\mu},q}(4\!-\!i)\varphi_{\tilde{\mu},q}(3\!-\!i)
\end{multline}
where $\bigl({\scriptstyle n\atop\scriptstyle i}\bigr)_q$ denotes $q$-binomial coefficient.
Taking averages $\langle...\rangle$ of the both sides of (\ref{recur2}),
after some algebra we find (recall that $z\equiv e^x$):
\begin{multline}\label{expr3}
\langle (a^\dag)^2 a^2\rangle \!=\! \frac{\varphi_{\tilde{\mu},q}(2)}{(z\!-\!q)(z\!-\!q^2)}
\biggl\{1 + \frac{(\varphi_{\tilde{\mu},q}(3)\!-\![3]_q)}{(z-q^3)(z-q^4)}\cdot\\
\shoveleft{\cdot \Bigl(z \!+\!q^2\!\!-\!\frac{q^2 [4]_q}{\varphi_{\tilde{\mu},q}(2)}\Bigr)\biggr\}
+ \frac{[N_{\rm max}\!+\!1]^{-1}_z}{(z\!-\!q)..(z\!-\!q^4)}
\bigl\{\varphi_{\tilde{\mu},q}(3)\varphi_{\tilde{\mu},q}(2)-}\\
\shoveleft{- \varphi_{\tilde{\mu},q}(N_{\rm max}\!+4)\varphi_{\tilde{\mu},q}(N_{\rm max}\!+3) + (z\!-\!q[4]_q) \cdot}\\
\cdot\bigl(\varphi_{\tilde{\mu},q}(2) \!-\! \varphi_{\tilde{\mu},q}(N_{\rm max}\!+\!3)
\varphi_{\tilde{\mu},q}(N_{\rm max}\!+\!2)\bigr)\bigr\}.
\end{multline}
The second part of this expression (beginning with $[N_{\rm max}\!+\!1]^{-1}_z$)
can be evaluated using the relations
\begin{equation}
\varphi_{\tilde{\mu},q}(N_{\rm max}+l) = -q [l-1]_q \frac{[N_{\rm max}+l]_q}{[N_{\rm max}]_{q^{-1}}},\ \ \
l\!=\!2,3,...,
\end{equation}
that yields
\begin{multline*}
\varphi_{\tilde{\mu},q}(l)\varphi_{\tilde{\mu},q}(l\!-\!1) - \varphi_{\tilde{\mu},q}(N_{\rm max}\!+\!l)
\varphi_{\tilde{\mu},q}(N_{\rm max}\!+\!l\!+\!1) =\\
\shoveleft{= -q^{l-1} [2]_q [l]_q [l\!-\!1]_q [N_{\rm max}\!+\!1]_{q^2}\cdot}\\
\cdot\bigl\{q^l(q\!-\!1) \!+\! (2q^{l+1}\!-\![2]_q) \tilde{\mu} \!+\! q(q[l]_q\!+\![l\!-\!1]_q) \tilde{\mu}^2\bigr\}.
\end{multline*}
After substituting these expressions in (\ref{expr3}) we obtain
\begin{multline}\label{<chi(N)>}
\langle (a^\dag)^2 a^2\rangle \!=\! \frac{\varphi_{\tilde{\mu},q}(2)}{(z\!-\!q)(z\!-\!q^2)} \biggl\{1 +
\frac{\delta\varphi_{\tilde{\mu},q}(3) \bigl(z\!+\!q^2 \!-\! \frac{q^2 [4]_q}{\varphi_{\tilde{\mu},q}(2)}\bigr)}{(z-q^3)(z-q^4)}\cdot\\
\cdot (1\!-\!R) - q^2 [2]_q R \frac{\Bigl(\frac{[2]_q [3]_q}{\varphi_{\tilde{\mu},q}(2)}\!-\!2q\!-\!1\Bigr) (z\!-\!q^2)
+ q^2 (q\!-\!1)^2}{(z-q^3)(z-q^4)} \biggr\}.
\end{multline}
Recall that expression (\ref{<chi(N)>}) is valid for the discrete case.
For the continuous case, take the limit $N_{\rm max}\to \infty$ to obtain
\begin{multline}\label{<(a)^2a^2>2}
\langle (a^\dag)^2 a^2 \rangle = \frac{\varphi_{\tilde{\mu},q}(2)}{(z\!-\!q)(z\!-\!q^2)}\cdot\\
\cdot\biggl\{1 + \frac{(\varphi_{\tilde{\mu},q}(3)\!-\![3]_q) \bigl(z \!-\! q^2 \bigl(\frac{[4]_q}{\varphi_{\tilde{\mu},q}(2)}\!-\!1\bigr)\bigr)}{(z-q^3)(z-q^4)}\biggr\}.
\end{multline}
Formulas~(\ref{<chi(N)>})-(\ref{<(a)^2a^2>2}) {\it constitute our second result}.

Now consider the limit cases:
\begin{align*}
&\text{If}\ \tilde{\mu} = N_{\rm max}^{-1}>0,\\
&\langle (a^\dag)^2 a^2 \rangle \mathop{\longrightarrow}\limits_{q\to 1} 2 (z^{\tilde{\mu}^{-1}+1}\!-\!1)^{-1}
(z\!-\!1)^{-4} \bigl\{(1\!-\!\tilde{\mu}) (z^{\tilde{\mu}^{-1}\!+3}\!-\!1)\\
&+ (6\tilde{\mu}^2\!-\!4\tilde{\mu}\!-\!2) (z^{\tilde{\mu}^{-1}\!+2}\!-\!z) +
(6\tilde{\mu}^2\!+\!5\tilde{\mu}\!+\!1) (z^{\tilde{\mu}^{-1}\!+1}\!-\!z^2)\bigr\};\vspace{1mm}\\
&\text{If}\ \tilde{\mu}<0,\\
&\langle (a^\dag)^2 a^2 \rangle \mathop{\longrightarrow}\limits_{q\to 1}
\frac{[2]_{\tilde{\mu}}}{(z\!-\!1)^2}
\Bigl\{1 + \frac{([3]_{\tilde{\mu}}\!-\!3) (z \!-\!4/[2]_{\tilde{\mu}}\!+\!1)}{(z-1)^2}\Bigr\}.
\end{align*}
Similarly to $\langle a^\dag a \rangle$, letting
$\tilde{\mu}\to 0$ in $\langle (a^\dag)^2 a^2 \rangle$ we find:
\begin{align}
&\text{If}\ |q|<1,\ \ \ \langle (a^\dag)^2 a^2 \rangle
\mathop{\longrightarrow}_{\tilde{\mu}\to 0} \frac{[2]_q}{(e^x-q)(e^x-q^2)};\\
&\text{If}\ q>1,\ \ \ \langle (a^\dag)^2 a^2 \rangle_{\tilde{\mu}\to 0} \simeq
\frac{[2]_q}{(z-q)(z-q^2)}-\nonumber\\
&-\theta(\tilde{\mu}) \frac{q^3 [2]_q (z\!-\!1)}{(z\!-\!q^2)(z\!-\!q^3)(z\!-\!q^4)}
\Bigl(\frac{\tilde{\mu}}{q\!-\!1}\Bigr)^{\frac{x}{\ln q}-2}\mathop{\longrightarrow}\limits_{\tilde{\mu}\to 0}\nonumber\\
&\mathop{\longrightarrow}\limits_{\tilde{\mu}\to 0}
\left\{
\begin{aligned}
&\frac{[2]_q}{(e^x\!-\!q)(e^x\!-\!q^2)},\ x>2\ln q;\\
&\infty,\ x<2\ln q.
\end{aligned}
\right.
\end{align}

\section{Intercept of two-particle correlation function}\label{sec:2nd-intercepts}

The substitution of~(\ref{<chi(N)>}) and~(\ref{<a^+a>2}) in (\ref{lambda_def})
leads us to the following resulting {\it expression for the intercept}:
\begin{multline}\label{lambda^(2)}
\lambda^{(2)} = -1 +\frac{\varphi_{\tilde{\mu},q}(2) (z-q)(z-q^2)}{\bigl(z\!-\!q^2\!+\!(1\!-\!R_{\tilde{\mu},q}(z)) \delta\varphi_{\tilde{\mu},q}(2)\bigr)^2 (z\!-\!q^3)(z\!-\!q^4)}\cdot\\
\Bigl\{ (z\!-\!q^3)(z\!-\!q^4) + (1\!-\!R_{\tilde{\mu},q}(z)) \delta\varphi_{\tilde{\mu},q}(3)
\Bigl(z\!+\!q^2 \!-\!{\textstyle\frac{q^2 [4]_q}{\varphi_{\tilde{\mu},q}(2)}}\Bigr) -\\
\!-\! q^2 [2]_q R_{\tilde{\mu},q}(z) \Bigl(\Bigl({\textstyle\frac{[2]_q [3]_q}{\varphi_{\tilde{\mu},q}(2)}}\!-\!2q\!-\!1\Bigr)
(z\!-\!q^2) \!+\! q^2 (q\!-\!1)^2\Bigr)\Bigr\}.
\end{multline}
In the continuous case this expression reduces to
\begin{multline}
\lambda^{(2)} = -1 +\frac{\varphi_{\tilde{\mu},q}(2) (z-q)(z-q^2)}{(z\!-\![3]_q\!+\!
\varphi_{\tilde{\mu},q}(2))^2 (z\!-\!q^3)(z\!-\!q^4)}\cdot\\
\cdot\!\Bigl\{(z\!-\!q^3)(z\!-\!q^4) + (\varphi_{\tilde{\mu},q}(3)\!-\![3]_q)
\Bigl(z\!+\!q^2 \!-\!{\textstyle\frac{q^2 [4]_q}{\varphi_{\tilde{\mu},q}(2)}}\Bigr)\Bigr\}.
\end{multline}
These expressions for $\lambda^{(2)}$ give {\it main result of the paper}.

Now, by substituting the obtained limits for $\langle a^\dag
a \rangle$ and $\langle (a^\dag)^2 a^2 \rangle$ in
(\ref{lambda_def}) we find the limits for the intercept:
\newline
\vspace{2mm}
If $q\to 1$ (and either $\tilde{\mu}>0$ or $\tilde{\mu}<0$),
\begin{align*}
&\lambda^{(2)}_{\tilde{\mu}>0} \mathop{\longrightarrow}\limits_{q\to 1}
\bigl[(1\!-\!2 \tilde{\mu}) (z^{2 \tilde{\mu}^{-1}\!+4}\!+\!1)\!+\!(12 \tilde{\mu}^2\!-\!4 \tilde{\mu}\!-\!2)
(z^{2 \tilde{\mu}^{-1}\!+3}\!+\!z)\\
&+\! (8 \tilde{\mu}^2\!+\!6 \tilde{\mu}\!+\!1) (z^{2 \tilde{\mu}^{-1}\!+2}\!+\!z^2) \!-\!
(12 \tilde{\mu}^2\!+\!12 \tilde{\mu}\!+\!6) z^{\tilde{\mu}^{-1}\!+1}(z^2\!+\!1)\\
&-\! (16 \tilde{\mu}^2\!-\!24 \tilde{\mu}\!-\!12) z^{\tilde{\mu}^{-1}\!+2}\bigr]/(z^{\tilde{\mu}^{-1}\!+2}
\!\!-\!1 \!-\! (2 \tilde{\mu}\!+\!1) (z^{\tilde{\mu}^{-1}\!+1}\!\!-\!z))^2\!,\\
&\lambda^{(2)}_{\tilde{\mu}<0} \mathop{\longrightarrow}\limits_{q\to 1} [2]_{\tilde{\mu}}
\frac{(z-1)^2 + ([3]_{\tilde{\mu}}\!-\!3)
(z \!-\! 4/[2]_{\tilde{\mu}}\!+\!1)}{(z-1-2\tilde{\mu})^2} - 1;
\end{align*}
If $\tilde{\mu}\to 0$ (and either $|q|\le 1$ or $q>1$),
\begin{align*}
&\lambda^{(2)}_{-1<q<1} \mathop{\longrightarrow}\limits_{\tilde{\mu}\to 0} q \frac{e^x-1}{e^x-q^2},\\
&\lambda^{(2)}_{q>1} \mathop{\simeq}\limits_{\tilde{\mu}\to 0}
\left\{
\begin{aligned}
&q \frac{e^x-1}{e^x-q^2},\ \ x>2\ln q,\\
&\frac{q^3 [2]_q (z\!-\!1) (z\!-\!q)^2}{(q^2\!-\!z)(q^3\!-\!z)(q^4\!-\!z)} \Bigl(\frac{\tilde{\mu}}{q\!-\!1}\Bigr)^{\frac{x}{\ln q}-\!2}
\rightarrow \infty,\\
&\qquad\qquad\qquad \ln q<x<2\ln q,\ \tilde{\mu}>0,\\
&\frac{q [2]_q (z\!-\!q)^2 (q^2\!-\!z)}{(z\!-\!1)(q^3\!-\!z)(q^4\!-\!z)} \Bigl(\frac{\tilde{\mu}}{q\!-\!1}\Bigr)^{-\frac{x}{\ln q}}
\rightarrow \infty, \\
&\qquad\qquad\qquad\qquad\qquad x<\ln q,\ \tilde{\mu}>0.
\end{aligned}
\right.
\end{align*}

Strictly speaking, the version of $\tilde{\mu},q$-deformed Bose gas
model (explored in this and preceding sections) in which we have derived
the one-, two-particle distribution functions and the intercept of two-particle
correlation function, is not identical to the $\tilde{\mu},q$-Bose gas model
of~\cite{GM_Virial,GM2014VirCoefs} and of Section \ref{sec:vir_exp}. Between the
two versions of $\tilde{\mu},q$-Bose gas there is a kind of ``duality relation'':
the base of this relation lies both in ({\it i}) the usage of the same
DSF (in the form of $\varphi_{\tilde{\mu},q}\bigl(z\frac{\rm d}{{\rm
d}z}\bigr)$ to deform thermodynamics, or in the form
$\varphi_{\tilde{\mu},q}(N)$ to calculate deformed distributions and
correlation intercepts), and in ({\it ii}) the required  %%following fact of
coincidence of one-particle deformed distributions. We mean the
distribution $n^{(\varphi)}_{\bf k}$ in $\varphi$-deformed model
defined like in Sec.~\ref{sec:vir_exp}
and~\cite{GM_Virial,GM2014VirCoefs} (with $\varphi$-deformed expression
for the total number of particles $N^{(\varphi)}$ and the corresponding
partition function) and recovered from $N^{(\varphi)} = \sum_{\bf k}
n^{(\varphi)}_{\bf k}$, on the one hand, and the distribution
$n^{(\tilde{\varphi})}_{\bf k}\equiv \langle\tilde{\varphi}(N_{\bf
k})\rangle$ defined by DSF $\tilde{\varphi}$ similarly
to~(\ref{<a^+a>2}), on the other hand. For these distributions to
coincide, the DSFs $\tilde{\varphi}$ and $\varphi$ should be properly
related. This ``duality'', in explicit terms and with concrete
examples, will be the subject of a separate paper. Here we only
mention that for the $\tilde{\mu}$-Bose gas model given by
$\varphi_{\tilde{\mu},q}(n)|_{q=1} = (1+\tilde{\mu})n - \tilde{\mu}
n^2$ which is the $q=1$ sector of the whole $\tilde{\mu},q$-Bose gas
model, the distributions and intercepts in the dual version are
calculated using the ``dual'' structure function
$\tilde{\varphi}_{\tilde{\mu},q}(n) = (1+\frac{\tilde{\mu}}{2})n -
\frac{\tilde{\mu}}{2} n^2$.

\begin{figure}[h]
\centering
\includegraphics[width = 1\linewidth]{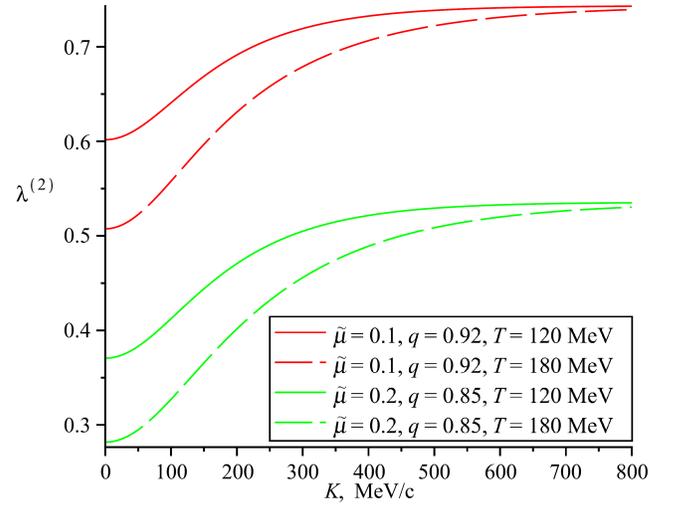}
\caption{Dependence of the two-pion intercept $\lambda^{(2)}(K)$ on
momentum $K$, for the values $\tilde{\mu}=0.1,\,0.2$,
$q=0.85,\,0.92$, and $T=120,\,180\,MeV$.} \label{fig2}
\end{figure}

\begin{figure}[h]
\centering
\includegraphics[width = 1\linewidth]{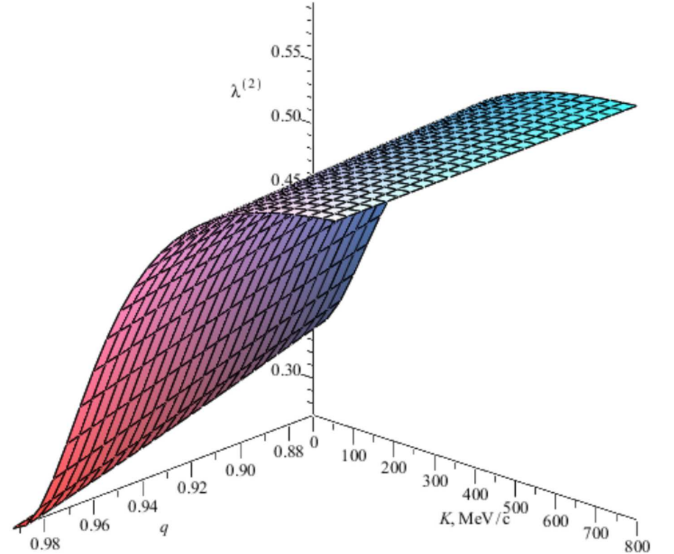}
\vspace{-5mm}
\caption{Intercept $\lambda^{(2)}(q,\lambda^{(2)}_{as};K)$ vs. momentum $K$ and
deformation parameter $q$ for temperature $T=180\, MeV$, at fixed
$\lambda^{(2)}_{as} = 0.6$ so that $\tilde{\mu} = \frac{1-0.6/q}{1+q}$.}
\label{fig4}
\end{figure}

The dependence $\lambda^{(2)}(K)$ on the momentum $K=|{\bf k}|$ for
some values of deformation parameters $\tilde{\mu}$, $q$ and
temperature $T$ is shown in Fig.~\ref{fig2}, where we take
$\hbar\omega = \sqrt{m^2+K^2}$, and for $m$ the $\pi$-meson mass (139.5
{\it MeV}).
In addition we give three-dimensional plot of the function
$\lambda^{(2)}(K,q)$ with fixed $\lambda^{(2)}_{as}$ in Fig.~\ref{fig4}.
%\vspace{-4mm}

\begin{figure}[h]
\centering
\includegraphics[width = 1\linewidth]{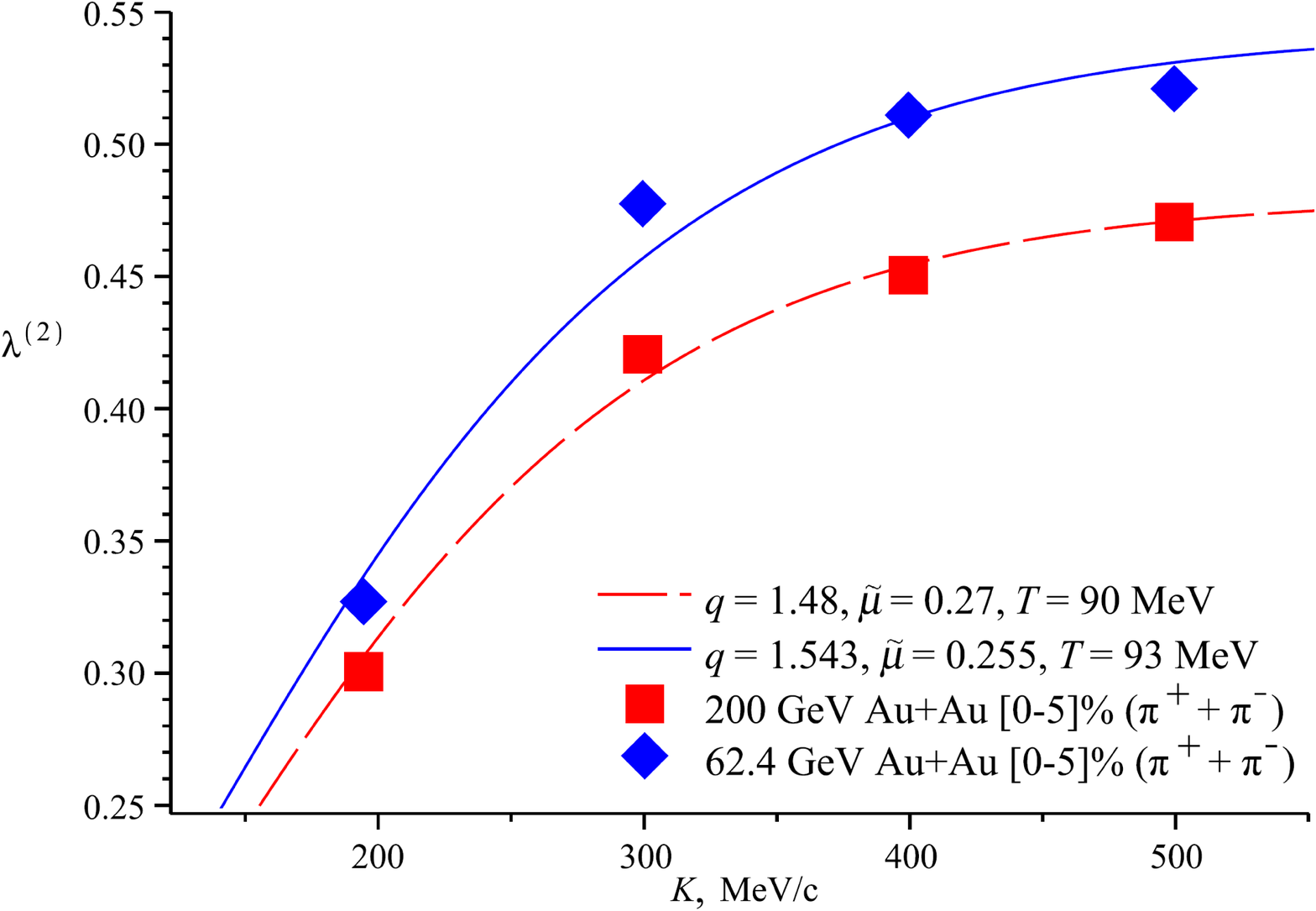}\vspace{2mm}\\
\includegraphics[width = 1\linewidth]{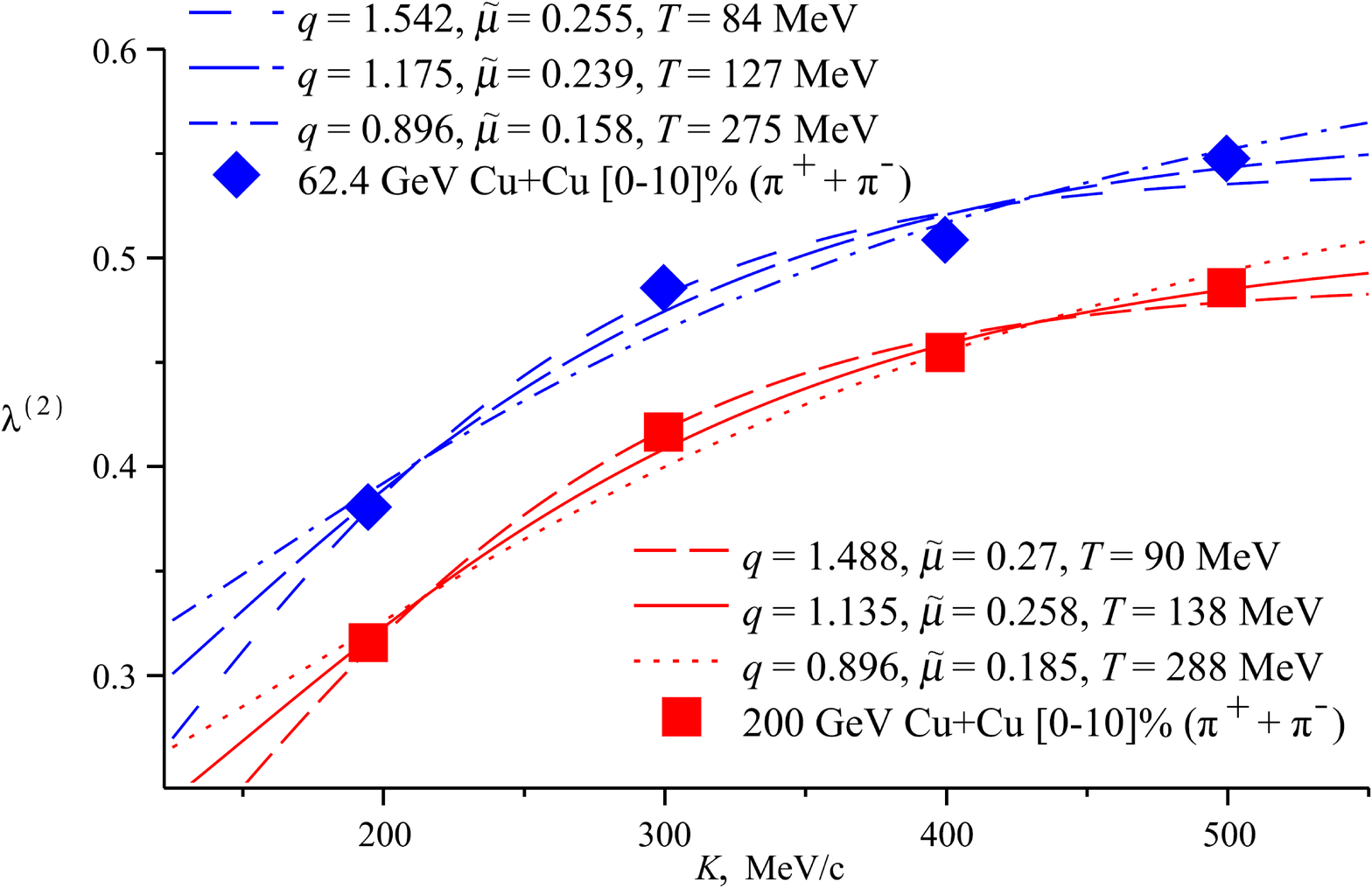}\\
\includegraphics[width = 1\linewidth]{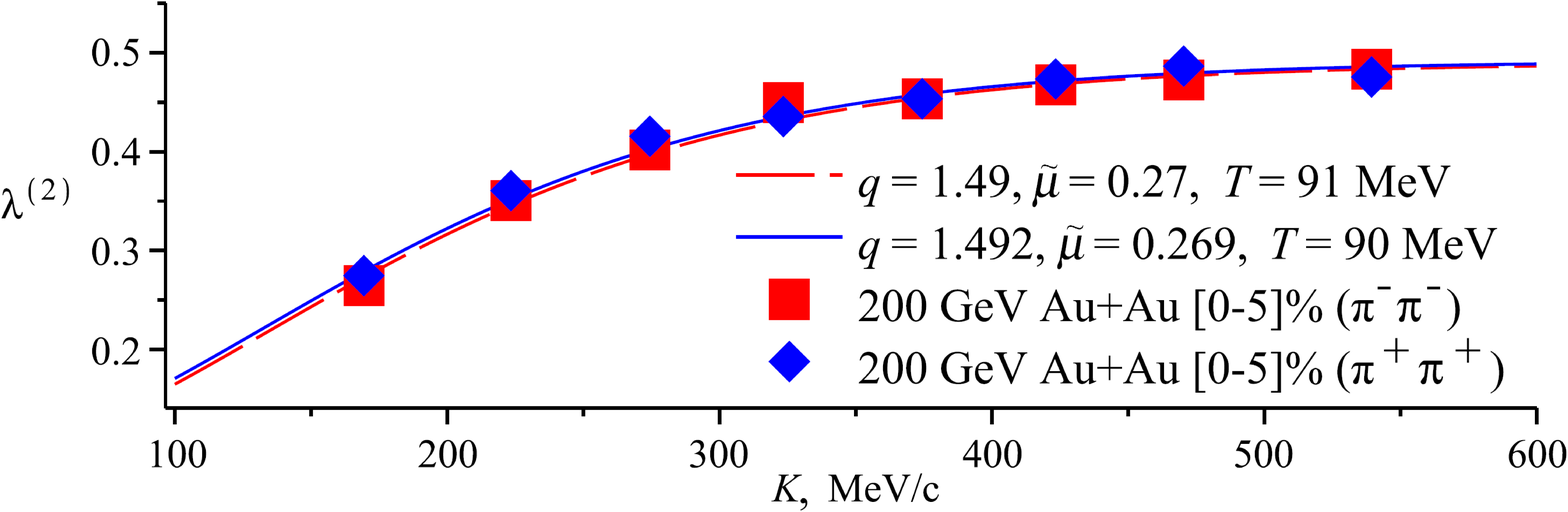}
\caption{Intercept $\lambda^{(2)}(K)$ vs. momentum $K$, for different values
of $q$, $\tilde{\mu}$ and $T$ chosen to fit experimental data. Exper. dots taken
from~\cite{STAR2005Pion,STAR_pion2009} are shown by boxes.}
\label{fig3}
\end{figure}

To confront our results with experimental data it is more convenient to
work (instead of the couple $\tilde{\mu},\,q$) with the asymptotic
value $\lambda^{(2)}_{as}$ of the intercept and one of the parameters
$\tilde{\mu}$, $q$, say $q$. The asymptotics is
\begin{equation}
\lambda^{(2)}_{as} = \varphi_{\tilde{\mu},q}(2)!-1 = q[1-\tilde{\mu}(1+q)].
\end{equation}
Expressing $\tilde{\mu}$ through $q$ and $\lambda^{(2)}_{as}$ as
$\tilde{\mu} = \frac{1-\lambda_{as}/q}{1+q}$ and substituting this
in (\ref{lambda^(2)}), we obtain the function $\lambda^{(2)}(q,\lambda_{as};K)$.
The corresponding plots for some values of deformation parameters $\tilde{\mu}$,
$q$ and temperature $T$ are presented in Fig.~\ref{fig3}. Therein, we also
place some experimental points for $\pi$-meson intercepts from RHIC/STAR,  %%in heavy ion collisions,
see~\cite{STAR2005Pion,STAR_pion2009}.
As seen, there is a qualitative agreement of theoretical curves with the data.
At first sight it is the $\pi$-meson compositeness effects which should be more
sensitive to the increase of collision energy of colliding ions. Therefore, under such a
condition within our interpretation, parameter $\tilde{\mu}$ somewhat more notably varies
than $q$ with the change of the collision energy for the same colliding ions.
So, besides an independent variation of $\tilde{\mu}$ and $q$ involved in
the extrapolation, the second panel of Fig.~\ref{fig3} contains also the two curves
with the same $q=0.896$ both for $62.4$ GeV and $200$ GeV Cu+Cu-collisions.
All the plots in Fig.~\ref{fig3} correspond to the discrete case. The analogous
fitting curves for the continuous case show somewhat worse agreement and are not shown.
As seen from the third panel of Fig.~\ref{fig3} the experimental dots for
$\pi^+$-meson intercept in $200$ GeV Au+Au collisions lie mainly slightly higher than
those for $\pi^-$-mesons. This is presumably explained by different effective Coulomb
interaction for $\pi^+$- and $\pi^-$-mesons, and can be associated with slightly differing
values (1.49 versus 1.492) of parameter $q$ of the corresponding extrapolating curves.
It is clear that more detailed experimental information is needed in order to
make more univocal conclusions about advantages of this model over others.
%----------------------------------------------------------------------------------------------------

\section{Conclusions and outlook}

In this work, within the $\tilde{\mu},q$-deformed Bose gas model
based on the deformation structure function (\ref{phi_mu_q}) taken
as operator function of the number operator, we have calculated both
one- and two-particle distribution functions from which obtained the
expression for momentum correlation function intercept. It should be
stressed that in this particular model, unlike other deformed models,
see~\cite{Anchishkin,Anchishkin_Transverse,Gavrilik_Sigma,GR_EPJA}
and some others, the deformation parameters $\tilde{\mu}$ and/or $q$
may take not only continuum values but also the discrete ones. Of course,
that can be presumably interpreted as inherited from the particles' compositeness
picture~\cite{GKM2,GKM,GM_Entang,GM_Ent(En)}.

The version of $\tilde{\mu},q$-deformed Bose gas model considered in
Sec.~\ref{sec:2nd-distr} and~\ref{sec:2nd-intercepts} of this paper,
though differs from its dual $\tilde{\mu},q$-deformed partner model
from~\cite{GM2014VirCoefs,GM_Virial} and Sec.~\ref{sec:vir_exp}
above, shares with it three things: the same form of DSF (as main
ingredient of any deformed model); the coinciding one-particle
distributions derived with differing, but strictly related (in a special way)
DSFs in the two models; the same impact, or goal, of effective description
of the two basic non-ideality properties of realistic Bose like
gases, mentioned in Introduction.  More detailed analysis of the
duality relation between the two partner $\tilde{\mu},q$-deformed
Bose gas models, with some other instances of dual pairs of
deformed models, will be the subject of forthcoming paper, including
the specifics of implications in the two dual approaches for the goal
of effective description.

What concerns application of the obtained $\tilde{\mu},q$-deformed two-particle
correlation function intercept for an effective description of the
observed non-Bose like behavior of two-pion correlation intercepts
observed in RHIC and LHC experiments
\cite{STAR2001pion,Bearden(CERN)2001,STAR2003HBTCorr,Aggarwal(WA98)2003,STAR2005Pion,STAR_pion2009,ALICE2014pion},
we applied the results obtained above and made some comparison which
shows a qualitative agreement.
%%%%%
Of course, for more detailed comparison and ultimate conclusion about
viability of the studied deformed Bose gas models, the knowledge of
(both formulas and data on) the 3rd order distribution function and respective
correlation intercept no doubt is desirable, also in view of the existing
characteristic function $r^{(3)}$ introduced in~\cite{Heinz_Zhang} (with some
deformed cases studied in~\cite{Gavrilik_Sigma,GR_EPJA}), which is a
special combination of the two- and three-particle correlation
function intercepts $\lambda^{(2)}$ and $\lambda^{(3)}$. The
particular results for 3rd order correlation function intercept
$\lambda^{(3)}$ and the function $r^{(3)}$ obtained for the
$\tilde{\mu},q$-deformed Bose gas, along with confronting them with available
experimental data for $\pi$-mesons (as quark-antiquark {\it composites}),
produced and registered in relativistic heavy-ion collisions, will
be presented elsewhere.

\section*{Acknowledgements}
\vspace{-1mm}
This work was partly supported by the Special Program of the Division of
Physics and Astronomy of National Academy of Sciences of Ukraine, by the Grant (Yu.A.M.)
for Young Scientists of National Academy of Sciences of Ukraine (No.~0113U004910), and by the Grant
(Yu.A.M.) of the President of Ukraine for Young Scientists (No.~0114U007147).
%=====================================================================================================

%\nocite{*}
\bibliographystyle{apsrev4-1}
\bibliography{references}

\end{document}